\documentclass[12pt, onecolumn,draftcls]{IEEEtran}

\usepackage{cite}
\usepackage{latexsym}
\usepackage{epsfig}
\usepackage{verbatim}
\usepackage{amssymb}

\usepackage{graphicx}
\usepackage{color}
\usepackage{epsfig}
\usepackage[cmex10]{amsmath}
\usepackage{stfloats}
\usepackage{subfigure}
\usepackage{flushend}
\usepackage{stfloats}
\usepackage{amsmath,amssymb,bm,amsfonts,bbm}

\newtheorem{mytheorem}{Theorem}

\begin{document}
\title{\huge{Outage Analysis of Energy Harvesting based Opportunistic Cooperative Communication Systems}}
\author{Yanju Gu
%\thanks{This work was supported by a Discovery Grant from the Natural Sciences and Engineering Research Council (NSERC) of Canada.}
%\thanks{Y. Gu and S. A\"issa are with the Institut National de la Recherche Scientifique (INRS-EMT), University of Quebec, Montreal, QC, Canada; Email: \{yanju.gu, aissa@emt.inrs.ca\}. }
}

\maketitle
\begin{abstract}
\noindent Wireless energy harvesting constitutes an effective way to prolong the lifetime of wireless networks.
In this paper, an opportunistic decode-and-forward cooperative communication system is investigated, where the energy-constrained relays harvest energy from the received information signal and the co-channel interference (CCI) signals and then use that harvested energy to forward the correctly decoded signal to the destination.
Different from conventional relaying system with constant energy supplier, the transmission power of the energy-constrained relay depends on the available energy that harvested, which is a random process.
Best relay selection that takes into account all the impacting ingredients on the received signal quality at the destination is deployed. The exact closed-form expression of the outage probability is derived, and the optimal value of the energy harvesting ratio for achieving minimum outage is numerically investigated.
In addition, the impacts of the CCI signals on the system's outage and diversity performances are analyzed. It is shown that the proposed relaying scheme can achieve full diversity order equal to the number of relays without the need of fixed power supplier.
\end{abstract}
%
%\begin{keywords}
%\noindent Co-channel interference, cooperative networks, decode-and-forward, energy harvesting, opportunistic relay selection, relaying.
%\end{keywords}
%
\section{Introduction}
Energy harvesting has become a strong solution for providing green energy and is widely adopted in green communications \cite{EnergyCooper,EH2,Gu2,DingMultiUser}.
The energy is usually harvested from solar and wind power and then transmitted to the power grid \cite{5357331} with the knowledge of power states \cite{powerconf, powertsp}.
In recent years, radio frequency (RF) based energy harvesting techniques \cite{Ada,EH1} have received more and more attention as this RF energy is already pervasive in the wireless communication network, such as those ambient RF from TV and cellular communications.
In cooperative communication networks, the intermediate relays are randomly deployed for forwarding signals to the destination. These devices like sensors are typically equipped with batteries with limited operation life.
Replacing batteries for such devices is not an easy task as they may be located in hostile environments.
Therefore, equipping relays with the capability of RF energy harvesting can prolong the lifetime of the network \cite{wang2016smart}.

The relays can harvest energy from the received RF signal which is a superposition of the desired signal and the CCI signals. 
The cooperative decode-and-forward (DF) relaying systems with single energy-constrained relay have been studied in \cite{GuICC14, Gu2}, where co-channel interference (CCI) signals as useful energy are taking into account.
In this paper, we take a step further by considering multi-relay cooperation, which can provide further spatial diversity.
In conventional DF multi-relaying systems with constant power supply for the relay \cite{IkkiBestRelay}, once the relay decodes the information correctly, the outage performance at the destination only depends on the channel quality between the relay and the destination.
However, it is much more complicated for the energy harvesting based relaying system, since the energy constrained relay is operating with the random energy that harvested from the received information and interference signals.
In this work, the exact closed-form expression of the outage probability for the energy-harvesting-based multi-relay DF cooperative system is derived, where best relay selection is deployed.
The optimal value of the energy harvesting ratio for achieving minimum outage is numerically investigated.
In addition, the impacts of the CCI signals on the system's outage and diversity performances are also analyzed.
\section{Energy-Harvesting Based Relaying}
\subsection{System and Channel Models}
 A cooperative DF relaying system is considered, where the source $S$ communicates with the destination $D$ through the help of multiple energy-constrained intermediate relaying nodes ($R_1, R_2, \cdots, R_L$). Each node is equipped with a single antenna and operates in the half-duplex mode in which the node cannot simultaneously transmit and receive signals in the same frequency band.
Both, the first hop (source-to-relay) and the second hop (relay-to-destination), experience independent Rayleigh fading with the complex channel fading gains given by $h_i \sim CN(0,\Omega_{h_i})$ and $g_i \sim CN(0,\Omega_{g_i})$, respectively. The channels follow the block-fading model in which the channel remains constant during the transmission of a block and varies independently from one block to another.
The channel state information is only available at the receiver.
As aforementioned, the system operates in the presence of external interferers. Specifically, we assume that there is an aggregate CCI signal affecting the $i^{\rm th}$ relay. The channel fading gain between the interferer and the $i^{\rm th}$ relay, denoted  $\beta_{i}$, is modeled as $\beta_{i}\sim CN(0,\Omega_{\beta_{i}})$. The desired channels and the interference channel are supposed to be independent from each other.

\subsection{Wireless Energy Harvesting at the Relay}
The power-splitting based protocol, where a portion of the received power is utilized for energy harvesting and the remaining power is used for information processing, is adopted at the relay node. $P$ is the received power of the signal at the relay and $\theta$, with $0\leq\theta\leq1$, is the fraction of power that the relay harvests from the received interference and information signal. The remaining power is $(1-\theta)P$, for information transmission from the relay to the destination.

In the first-hop phase, the source $S$ transmits signal $s$ with power $P_{_S}$ to the relays. We consider a pessimistic case in which power splitting only reduces the signal power, but not to the noise power, which can provide a lower bound for relaying networks in practice. Accordingly, the received signal at the $i^{\rm th}$ relay for information detection is given by
\begin{equation} \label{Signal.P}
y_{_{SR_i}} = \sqrt {(1-\theta)P_{_S}}h_is + \sqrt{(1-\theta)P_{i}} \beta_{i}s_{i} + n_{_{R_i}},
\end{equation}
where $s_i$ and $P_i$ denote the signal and the corresponding power, respectively, from the interferer at the $i^{\rm th}$ relay, and $n_{_{R_i}}$ is the additive white Gaussian noise (AWGN) at the $i^{\rm th}$ relay with zero mean and variance $\sigma^2_R$.

According to (\ref{Signal.P}), the received signal-to-interference-plus-noise ratio (SINR) at the relay is given by
\begin{align}
\label{SINR.P}
\gamma _{_{SR_i}}
= \frac{(1-\theta)P_{_S}|h_i|^2}{\sigma_{_R}^2 + (1-\theta)P_{i}|\beta_{i}|^2}
= \displaystyle\frac{\gamma_{h_i}}{1 + I_i},
\end{align}
where ${\gamma}_{h_i} \triangleq\frac{(1-\theta)P_{_S}}{\sigma_{_R}^2}|h_i|^2$ and ${I}_i \triangleq \frac{(1-\theta)P_i}{\sigma^2_R}|\beta_i|^2$.
The relay harvests energy from the received information signal and the interference signal for a duration of $T/2$ at each block, and thus, the harvested energy at the $i^{\rm th}$ relay is obtained as
\begin{equation} \label{Harvest.P}
E_{_{H_i}} = \eta{\theta}\bigg(P_{_S} |h_i|^2 + P_{i}|\beta_{i}|^2\bigg)T/2,
\end{equation}
where $\eta$ is the energy conversion efficiency of the relay with value varying from $0$ to $1$ depending upon the harvesting circuitry.
It is assumed that there is no additional device like an energy buffer to store the
harvested energy (Harvest-Use) which can decrease the complexity of the energy harvesting nodes, so all the energy collected during the harvesting phase is consumed by the relay.
Since the processing power required by the transmit/receive circuitry at the relay is generally negligible compared to the power used for signal transmission \cite{TSplit, Neg}, here we suppose that all the energy harvested from the received signal is consumed by the relay for forwarding the information to the destination node.
The transmission power of the $i^{\rm th}$ relay is then given by
\begin{align} \label{PR.P}
P_{_{R_i}}
= \frac{{E_{_{H_i}}}}{T/2}
= \frac{{\eta \theta \sigma _R^2}}{1 - \theta }\left({\gamma}_{h_i}+{I}_i\right).
\end{align}
\subsection{Selective Transmission}
Let $\mathcal{L}=\left\{1,2,\ldots,L\right\}$ denote the set of all the $L$ relays and $\mathcal{S}$ represents the decoding subset consisted of those relays that are able to decode the source message. That is,
\begin{align} \label{S set}
\mathcal{S}
&\triangleq\left\{i\in\mathcal{L}: \gamma_{_{SR_i}}\geq \gamma_{_{th}} \right\},
\end{align}
where $\gamma_{_\mathrm{th}}$ is a pre-defined threshold.

In the second-hop phase, only a single relay among the relays belonging to subset $\mathcal{S}$ is allowed to transmit the signal. More specifically, from that decoding subset, the relay with the maximum signal-to-noise ratio (SNR) retransmits the information. That is, $\hat i =\arg\max_{i\in \mathcal{S}}\{\gamma_{_{R_iD}}\}$. Such technique is referred to as opportunistic relay selection.
Note that the relay selection process require centralized processing and how to extend it to distributed processing \cite{duJMLR, pairwise, informationmatrix} is still an open problem. Therefore, the received SNR at the destination, given set $\mathcal{S}$, can be written as
\begin{align}
\gamma_{_{R_{\hat i}D}}
=\frac{P_{_{R_{\hat i}}}|g_{\hat i}|^2}{\sigma_{_D}^2}
=\underbrace{\frac{\eta\theta }{( 1 - \theta) } \frac{\sigma_{_R}^2}{\sigma_{_D}^2 }
{|g_{\hat i}|^2}}_{\triangleq W_{\hat i}}({\gamma}_{h_{\hat i}}+{I}_{\hat i}),\label{SNR_D_BS}
\end{align}
where ${\sigma_{_D}^2 }$ is the variance of AWGN at the destination. The defined random variable $W_{\hat i}$ follows the same distribution as of $|g_{\hat i}|^2$.

Note that, in contrast to traditional DF relaying system with no rechargeable nodes, the transmission power $P_{_{R_i}}$ at the relay in the energy harvesting system is not a constant but a random variable, which depends on the replenished energy from the interference and information signals.
Therefore, the distribution of the received SNR at the destination is determined not only by the distribution of the relay-to-destination channel power gain $|g_{\hat i}|^2$, but also by the distribution of the information and interference signal power, i.e., ${\gamma}_{h_{\hat i}}$ and ${I}_{\hat i}$.

\section{Performance Evaluation}
\subsection{Outage Analysis}
%
%The exact closed-form expression of the outage probability of a opportunistic DF cooperative communication system, where the energy-constrained relays harvest energy from the received information signal and the interference signal, is now derived.

As an important performance measure of wireless systems, outage probability is defined as the probability that the instantaneous output SNR falls below a pre-defined threshold $\gamma_{_\mathrm{th}}$.
Mathematically speaking, $P_{\rm out}(\gamma_{_\mathrm{th}})={\rm Pr}\left(\gamma < \gamma_{_\mathrm{th}}\right)$.
This SNR threshold guarantees the minimum QoS requirement of the destination users.
In the DF relaying system under study, the outage probability at the destination, given set $\mathcal{S}$, is expressed as
\begin{align}
&\Pr \{ \gamma_{_{R_{\hat i}D}} < \gamma_{_\mathrm{th}}|\left| \mathcal{S} \right| = l\} \nonumber \\
=& \left[ \Pr \{ \gamma_{_{R_iD}} < \gamma_{_\mathrm{th}}\left| \gamma_{_{S{R_i}}} \ge \gamma_{_\mathrm{th}} \right.\}  \right]^l \label{c1} \\
=& \left[ \frac{\Pr \{ \gamma_{_{R_iD}} < \gamma_{_\mathrm{th}},\gamma_{_{S{R_i}}} \ge \gamma_{_\mathrm{th}}\}}{\Pr \{ \gamma_{_{S{R_i}}} \ge \gamma_{_\mathrm{th}}\} } \right]^l ,  \label{c2}
\end{align}
where $\left| \mathcal{S} \right|$ denotes the number of relays in the decoding set $\mathcal{S}$. Note that we assume a homogenous scenario where the desired channels as well as the interference channels are independent and identically distributed (i.i.d.), i.e., $\Omega_{h_i}=\Omega_{h}$, $\Omega_{g_i}=\Omega_{g}$, $\Omega_{\beta_i}=\Omega_{\beta}$, $P_i=P_I$ for $i = 1,2,\ldots,L$, and hence, the outage probability is independent of the combination of relays forming the decoding subset. Expression (\ref{c1}) is obtained by using the first principle of order statistics \cite{du2013network}. Since $ \gamma_{_{R_iD}}$ and $ \gamma_{_{SR_i}}$ are not known to be independent, this conditional probability is rewritten as (\ref{c2}), in which, the probability in the denominator is given by
\begin{align}\label{FirstPro}
\Pr \{ \gamma_{_{S{R_i}}} \ge \gamma_{_\mathrm{th}}\}
=& 1 - \Pr \{ \gamma_{_{S{R_i}}} < \gamma_{_\mathrm{th}}\}  \nonumber \\
=& \left( {1 + \frac{{\bar \gamma}_{_{\beta}}}{{{\bar \gamma }_{h}}}\gamma_{_\mathrm{th}}} \right)^{ - 1}\exp \left( { - \frac{\gamma_{_\mathrm{th}}}{{{\bar \gamma }_{h}}}} \right),\ \ \
\end{align}
where ${\bar\gamma}_{h} = \frac{(1-\theta)P_{_S}}{\sigma_R^2}\Omega_{h}$ and ${\bar\gamma}_{\beta}= \frac{(1-\theta)P_I}{\sigma_R^2}\Omega_{\beta}$.
The calculation of the probability in the numerator of (\ref{c2}) involves three random variables, ${\gamma}_{h_{i}}$, ${\gamma}_{g_{i}}$ and ${I}_{i}$; therefore, it is hard to get the result directly. By taking into account the physical meaning of this joint probability, it can be divided into two parts to simplify the derivation, that is,
\begin{align}\label{Joint}
&\Pr \{ \gamma_{_{R_iD}} < \gamma_{_\mathrm{th}},\gamma_{_{S{R_i}}} \ge \gamma_{_\mathrm{th}}\}\nonumber\\
=& \Pr \big \{ \gamma_{_{R_iD}}\mathbbm{1}_{\mathcal{C}} < {\gamma_{_\mathrm{th}}}\big \}-\Pr \{ \gamma_{_{S{R_i}}} < \gamma_{_\mathrm{th}}\},
\end{align}
where $\mathbbm{1}_{\mathcal{C}}$ is the indicator random variable for the set
$\mathcal{C}=\{\gamma_{_{SR_i}} \geq \gamma_{_\mathrm{th}}\}$, i.e.,
$\mathbbm{1}_{\mathcal{C}}=1$ if $\gamma_{_{SR_i}} \geq \gamma_{_\mathrm{th}}$, otherwise, $\mathbbm{1}_{\mathcal{C}}=0$.

\begin{mytheorem}\label{PrPower}
The first probability in (\ref{Joint}) is given by
\begin{align}\label{CDF_D}
\Pr \big \{ \gamma_{_{R_iD}}\mathbbm{1}_{\mathcal{C}} < {\gamma_{_\mathrm{th}}}\big \}
=& 1 - \frac{1}{{\bar \gamma }_h-{\bar \gamma }_{_\beta} } \bigg[{{\bar \gamma }_h}\Gamma\Big(1,\frac{\gamma_{_\mathrm{th}}}{\bar \gamma_h};\frac{\gamma_{_\mathrm{th}}}{{\bar \gamma }_h{\bar \gamma }_g}\Big)-b^{-1} \nonumber \\
 & \times\exp \left( {\frac{{{a}\gamma_{_\mathrm{th}}}}{{1 + \gamma_{_\mathrm{th}}}}} \right)
\!\Gamma\Big(1,b\gamma_{_\mathrm{th}};\frac{b\gamma_{_\mathrm{th}}}{\bar \gamma_g}\Big)\bigg],
\end{align}
where $a \triangleq \frac{{\rm{1}}}{{\bar \gamma }_{_\beta}} - \frac{{\rm{1}}}{{{\bar \gamma }_h}}$, ${b} \triangleq \frac{1}{{{{\bar \gamma }_h}}} + \frac{{a}}{{1 + \gamma_{_\mathrm{th}}}}$ and ${\bar \gamma }_g = \frac{\eta \theta}{1 - \theta } \frac{\sigma_R^2}{\sigma_D^2}\Omega_g$.
$\Gamma(a,x;b)$ is the generalized incomplete Gamma function defined by $\Gamma(a,x;b) \triangleq \int_{x}^{\infty}t^{a-1}\exp(-t-bt^{-1})d t$.
\end{mytheorem}
\begin{IEEEproof}
%See Appendix \ref{a-A}.
From (\ref{SNR_D_BS}), we get $\gamma_{_{R_iD}}\mathbbm{1}_{\mathcal{C}}= W_i({\gamma}_{h_i}+{I}_{i})\mathbbm{1}_{\mathcal{C}}$.
Define $Z \triangleq (\gamma_{h_i} + I_i)\mathbbm{1}_{\mathcal{C}}$,
then the cumulative distribution function (CDF) of $Z$ is given by
\begin{align}\label{CDFZ}
F_Z(z)
= \Pr\{Z<z\}
= \int\int_{{x,y}\in \mathcal{A}}f_{\gamma_{h_i},I_i}(x,y) \mathrm{d}x \mathrm{d}y,
\end{align}
where the set $\mathcal{A}={\left\{ {x + y < z,{\kern 1pt} \frac{x}{{1 + y}} > \gamma_{_\mathrm{th}}}, x\geq0, y\geq0 \right\}}$.
After some set manipulations, we have
$\mathcal{A}\neq \emptyset $ if and only if $ z> \gamma_{_\mathrm{th}}$.
Since $\gamma_{h_i}$ and $I_i$ are independent, we get the joint distribution $f_{\gamma_{h_i},I_i}(x,y) = f_{\gamma_{h_i}}(x)f_{I_i}(y)$.
Then, after some straightforward algebraic derivations, we obtain
\begin{equation}\label{Fz}
F_Z(z) =
\mathbbm{1}_{\mathcal{Z}}
\int_0^{ \frac{z - \gamma_{_\mathrm{th}}}{1 + \gamma_{_\mathrm{th}}}}
\int_{(1+y)\gamma_{_\mathrm{th}}}^{z - y} {f_{\gamma_{h_i}}(x)f_{I_i}(y)} \mathrm{d}x \mathrm{d}y.
\end{equation}
Both $\gamma_{h_i}$ and $I_i$ are of exponential distributions. Integrate with respect to $x$ and $y$ yielding the CDF of $Z$, where \cite[Eq.(3.351.1)]{Gradshteyn} was used. Then the probability density function (PDF) of $Z$, $f_Z(z)$, follows directly from differentiating $F_Z(z)$ with respect to $z$, and is given by
\begin{align}
f_Z(z)
=\mathbbm{1}_{\mathcal{Z}}
\frac{1}{{\bar \gamma }_h-{\bar \gamma }_{_\beta}}\exp (-\frac{z}{{\bar \gamma }_h})
\left[ 1\!-\! \exp \left( { - a\frac{z - \gamma_{_\mathrm{th}}}{1 + \gamma_{_\mathrm{th}}}} \right) \right],
\label{fZ3}
\end{align}
where $a \triangleq \frac{{\rm{1}}}{{\bar \gamma }_{_\beta}} - \frac{{\rm{1}}}{{{\bar \gamma }_h}}$ and $\mathbbm{1}_{\mathcal{Z}}$ is the indicator random variable for the set
$\mathcal{Z}=\{z> \gamma_{_\mathrm{th}}\}$, i.e.,
$\mathbbm{1}_{\mathcal{Z}}=1$ if $ z > \gamma_{_\mathrm{th}}$, otherwise, $\mathbbm{1}_{\mathcal{Z}}=0$.
Finally, we have
\begin{align}\label{Frd}
\Pr \big \{ \gamma_{_{R_iD}}\mathbbm{1}_{\mathcal{C}} < {\gamma_{_\mathrm{th}}}\big \}
 %&= \Pr \{W_iZ < \gamma_{_\mathrm{th}}\} \nonumber \\
 =& \mathbb{E}_Z \left\{ {1 - \exp \left( - \frac{\gamma_{_\mathrm{th}}}{{{{\bar \gamma }_g}Z}} \right)} \right\} \nonumber \\
 =& 1 - \int_0^\infty  {\exp \left( { - \frac{\gamma_{_\mathrm{th}}}{{{{\bar \gamma }_g}z}}} \right){f_Z}} \left( z \right)dz \nonumber \\
 =& 1 - \frac{1}{{\bar \gamma }_h-{\bar \gamma }_{_\beta} } \bigg[{{\bar \gamma }_h}\Gamma\Big(1,\frac{\gamma_{_\mathrm{th}}}{\bar \gamma_h};\frac{\gamma_{_\mathrm{th}}}{{\bar \gamma }_h{\bar \gamma }_g}\Big)-b^{-1} \nonumber \\
 & \times\exp \left( {\frac{{{a}\gamma_{_\mathrm{th}}}}{{1 + \gamma_{_\mathrm{th}}}}} \right)
\Gamma\Big(1,b\gamma_{_\mathrm{th}};\frac{b\gamma_{_\mathrm{th}}}{\bar \gamma_g}\Big)\bigg],
\end{align}
where ${b} \triangleq \frac{1}{{{{\bar \gamma }_h}}} + \frac{{a}}{{1 + \gamma_{_\mathrm{th}}}}$ and
$\Gamma(a,x;b)$ is the generalized incomplete Gamma function defined by $\Gamma(a,x;b) \triangleq \int_{x}^{\infty}t^{a-1}\exp(-t-bt^{-1})d t$, which completes the proof.
\end{IEEEproof}

According to the theorem of total probability, the unconditional outage probability at the destination, $P_{\mathrm{out}}\left( \gamma_{_\mathrm{th}} \right)$, is expanded as
\begin{align}\label{Outage}
P_{\mathrm{out}}\left( \gamma_{_\mathrm{th}} \right)
=&\sum_{l = 0}^{L}\Pr\{\left| \mathcal{S} \right| = l\} \Pr\{ \gamma_{_{R_{\hat i}D}} < \gamma_{_\mathrm{th}}|\left| \mathcal{S} \right| = l\} \nonumber\\
=&\left[ \Pr \{ \gamma_{_{S{R_i}}} < \gamma_{_\mathrm{th}} \} \right]^L  \nonumber \\
 & \times \sum_{l = 0}^{L}\binom{L}{l} \left[ \frac{\Pr \big \{ \gamma_{_{R_iD}}\mathbbm{1}_{\mathcal{C}} < {\gamma_{_\mathrm{th}}}\big \}}{\Pr \{ \gamma_{_{S{R_i}}} < \gamma_{_\mathrm{th}}\} } -1\right]^l,
\end{align}
where the following expression is used to obtain (\ref{Outage}),
\begin{align}
&\Pr\{\left| \mathcal{S} \right| = l\}\nonumber \\
=&\binom{L}{l} \left[ \Pr \{ \gamma_{_{S{R_i}}} < \gamma_{_\mathrm{th}} \} \right]^{L-l}
\left[ \Pr \{ \gamma_{_{S{R_i}}} \geq \gamma_{_\mathrm{th}} \} \right]^l.
\end{align}
\subsection{Numerical Results and Discussion}
Numerical examples are presented and corroborated by simulation results to examine the outage of the DF cooperative communication system, where the energy-constrained relays harvest energy from the received information signal and the CCI signals.
Hereafter, and unless stated otherwise, the threshold $\gamma_{_\mathrm{th}}$ is set to $5\mathrm{dB}$ and the energy conversion efficiency $\eta$ is set to $1$. To better evaluate the effect of the interference on the system's outage, we define $\frac{P_{_S}\Omega_h}{P_I\Omega_{\beta}}$ as the average signal-to-interference ratio (SIR) and $\frac{P_I\Omega_{\beta}}{\sigma_R^2}$ as the average interference-to-noise ratio (INR) at the first hop.

\begin{figure}[!ht]
%\vspace*{-0.1in}
\centering
\epsfig{figure=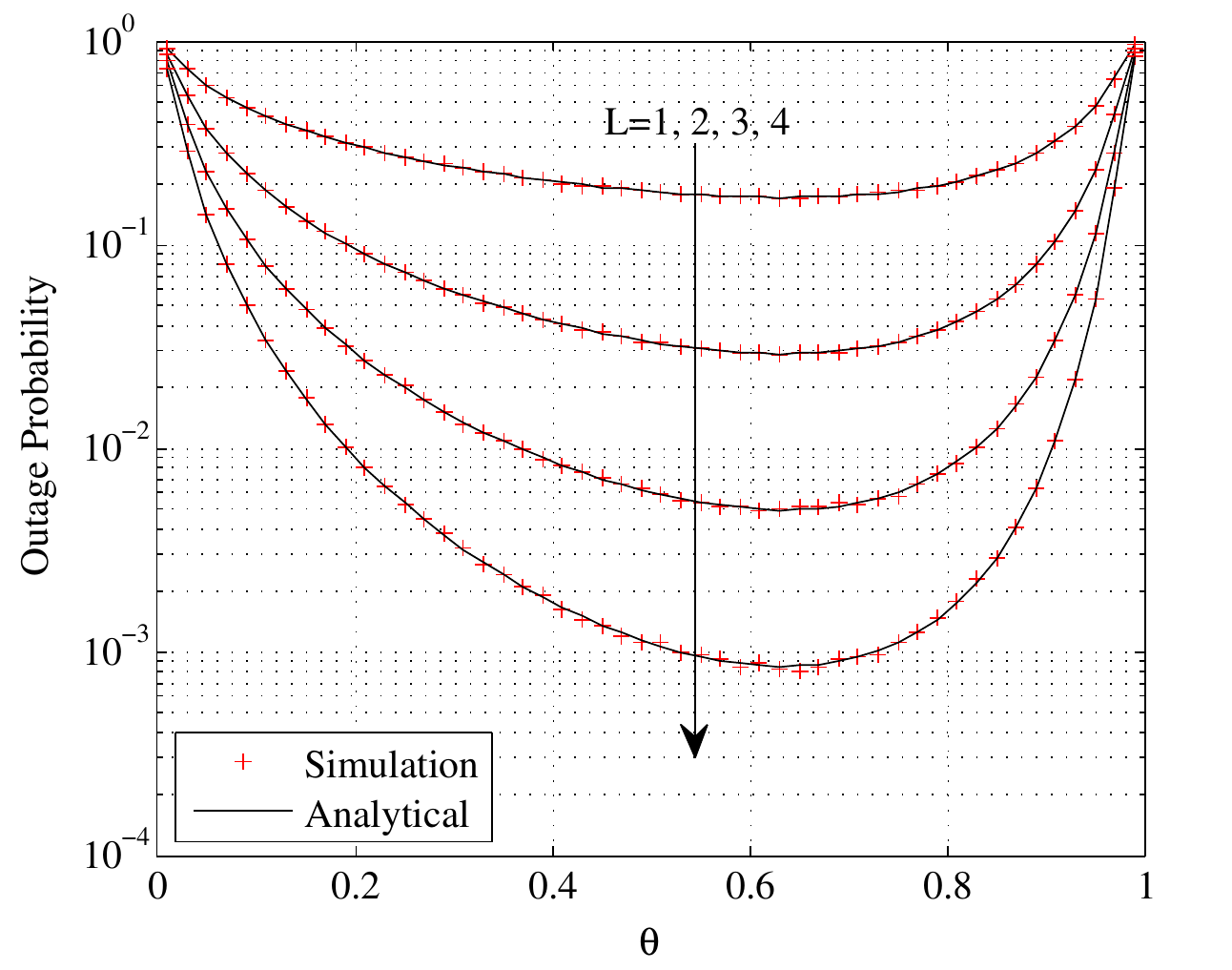, width=2.7in,height=2in}
%\vspace*{-0.05in}
\caption{Outage probability versus the energy harvesting ratio $\theta$ for different number of relays, where the first-hop average SNR is $20\mathrm{dB}$ and there is no interference at the relays.}
\label{Fig2}
\end{figure}

Figure \ref{Fig2} shows the outage probability versus the energy harvesting ratio $\theta$ for different number of relays, where the first-hop average SNR is $20\mathrm{dB}$, and there is no interference affecting the relay.
It is observed that the analytical results of (\ref{Outage}) match perfectly the simulation results.
As the number of available relays increases, the outage performance of the energy harvesting based relaying system improves.
The convex feature of the curves is due to the fact that the energy harvested for the second-hop transmission increases with increasing $\theta$, which effectively decreases the outage of the second hop and, accordingly, improves the outage performance of the system.
Meanwhile, as $\theta$ increases, more power is harvested for information transmission and less power is left for information decoding which increases the outage of the first hop and reduces the number of reliable relays for relay selection, therefore, the outage probability first drops down until reaching a minimum and then increases up.

\begin{figure}[!ht]
%\vspace*{-0.1in}
\centering
\epsfig{figure=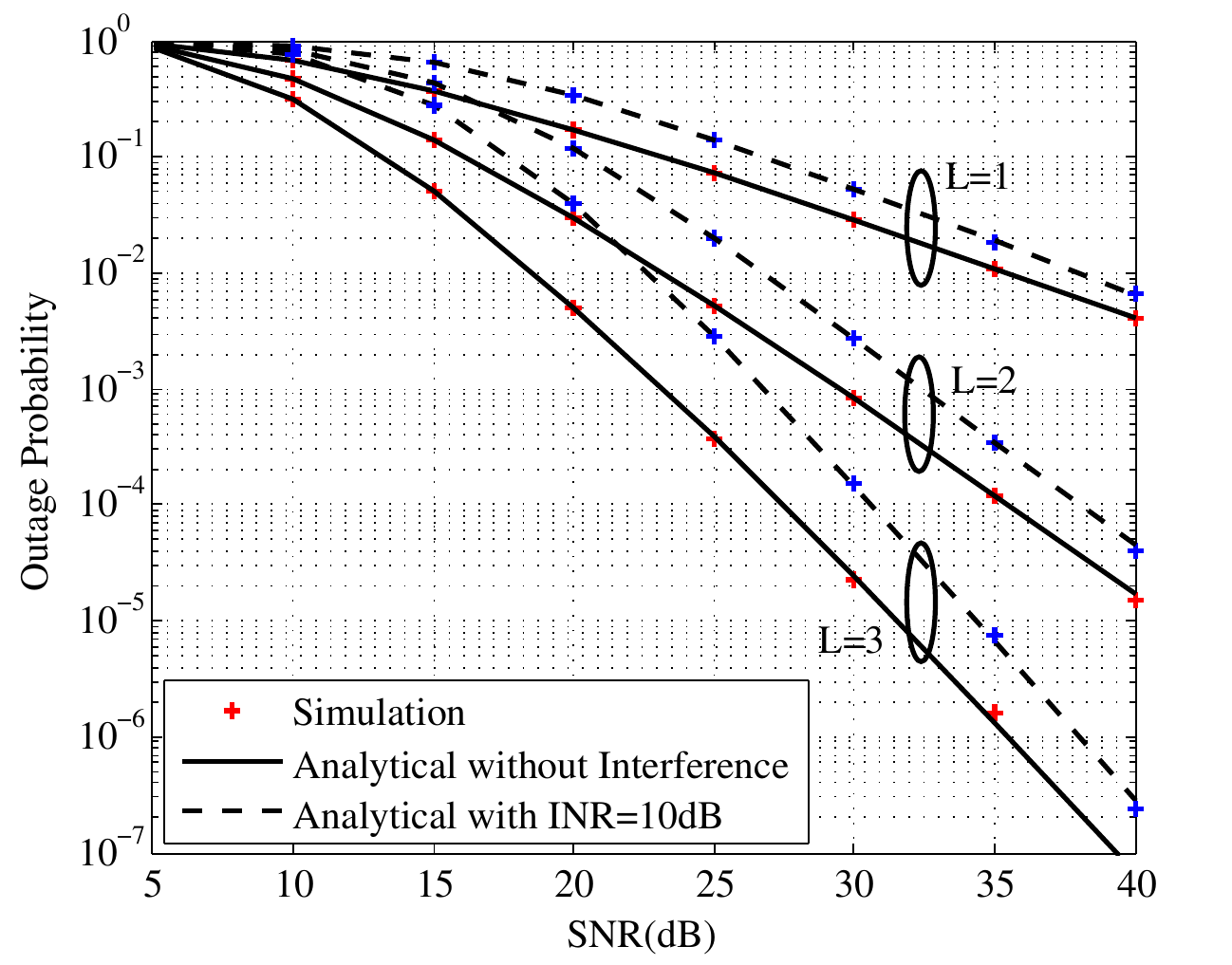, width=2.7in,height=2in}
%\vspace*{-0.05in}
\caption{Outage probability versus the first-hop average SNR for different number of relays, where the first-hop average INR is $10\mathrm{dB}$ and the energy harvesting ratio $\theta$ is set to $0.6$.}
\label{Fig3}
\end{figure}

\begin{figure}[!ht]
%\vspace*{-0.1in}
\centering
\epsfig{figure=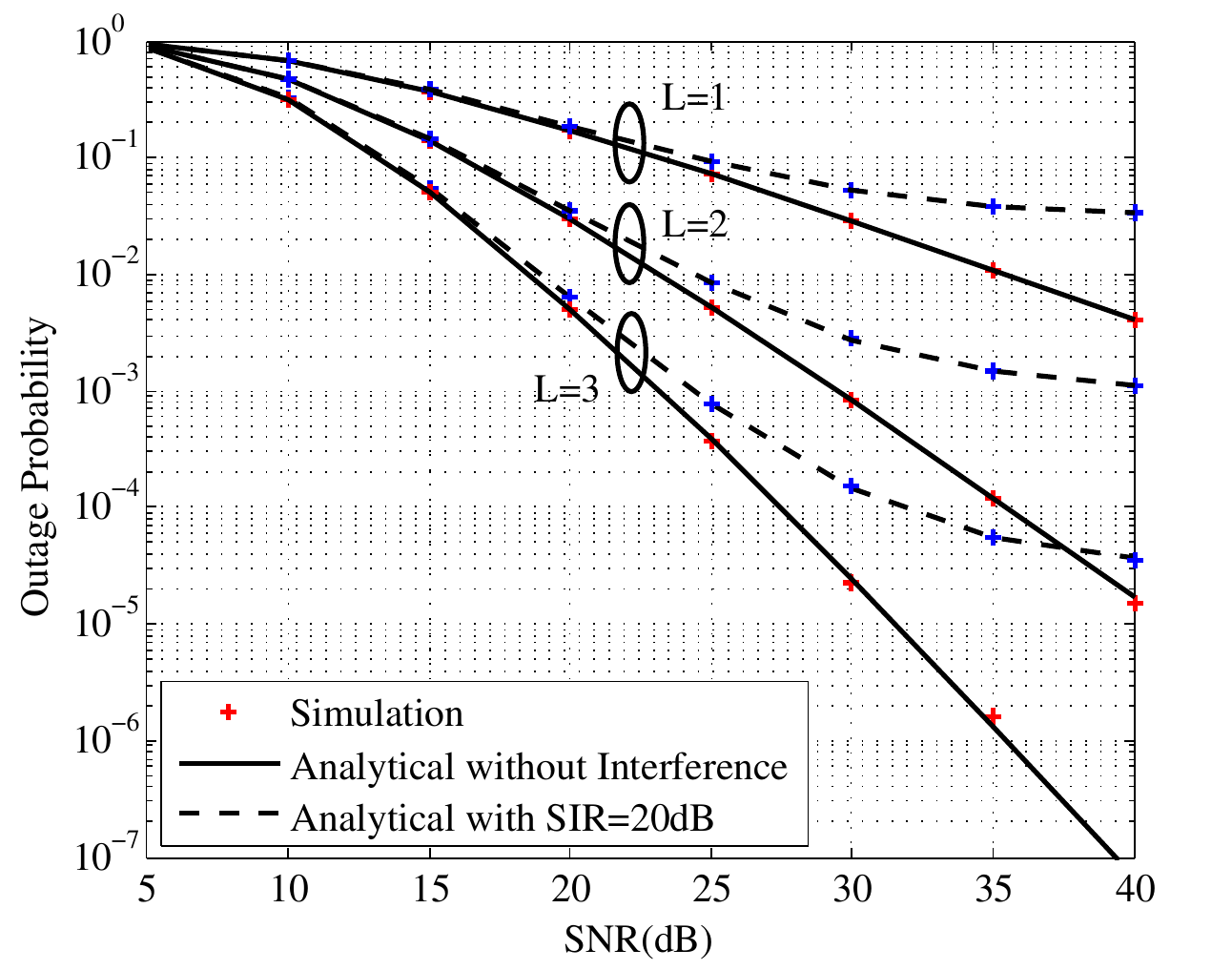, width=2.7in,height=2in}
%\vspace*{-0.05in}
\caption{Outage probability versus the first-hop average SNR for different number of relays, where the first-hop average SIR is $20\mathrm{dB}$ and the energy harvesting ratio $\theta$ is set to $0.6$.}
\label{Fig4}
\end{figure}

In order to clearly demonstrate the diversity gain of the system and the impacts of CCI signals on the system's performance, Figs. \ref{Fig3} and \ref{Fig4} illustrate the outage probability of the system versus the first-hop average SNR for different number of relays under two different situations where the first-hop average INR is $10\mathrm{dB}$ and the first-hop average SIR is $20\mathrm{dB}$, respectively. From Fig. \ref{Fig2}, it is clearly seen that the minimum outage is achieved when the energy harvesting ratio $\theta$ is of the value around $0.6$, so the ratio $\theta$ is set to $0.6$ here for performance evaluation.
It is shown from each plot in Figs. \ref{Fig3} and \ref{Fig4} that, even though the CCI signals can be utilized as useful energy during the energy harvesting phase, the existence of interference at the relay during the information decoding phase still limits the outage performance. It is also shown that when there is no interference affecting the relay, the spatial diversity order, which indicates the decreasing speed of the outage with respect to SNR, increases as the number of available relays increases. In Fig. \ref{Fig3}, it is seen that the diversity order remains the same when the interference keeps constant, while in Fig. \ref{Fig4}, the diversity order is reduced to zero and error floors appear in the high-SNR regime due to the constant interference to signal ratio.
\section{Conclusions}
In this paper, an opportunistic decode-and-forward (DF) cooperative communication system was studied, where the relays are energy-constrained and need to replenish energy from the received information signal and the co-channel interference (CCI) signals, and then use that harvested energy to forward the correctly decoded signal to the destination.
Different from traditional DF relaying system with no rechargeable nodes, the transmission power of the energy constrained relay is not a constant anymore but a random variable depending on the variation of available energy harvested from the received information and CCI signals at the relay. In order to better evaluate the system performance, the exact closed-form expression of the outage probability was derived. The optimal value of the energy harvesting ratio for achieving minimum outage was numerically investigated. The proposed relaying scheme can achieve full diversity order equal to the number of relays without the need of fixed power supply.
By applying some interference cancelation schemes at the information decoder, the system performance can be further improved.

%
%%%%%%%%%%%%%%%%%%%%%%%%%
%  References
%%%%%%%%%%%%%%%%%%%%%%%%
%\bibliographystyle{IEEEtran}
%\bibliography{IEEEabrv,EH_Ref}

\end{document}